\documentclass[aps,prd,11pt]{revtex4}
\usepackage[colorlinks=true, pdfstartview=FitV, linkcolor=blue, citecolor=red, urlcolor=magenta]{hyperref}
\usepackage{graphicx,color}
\usepackage{latexsym}
\usepackage{amsmath}
\usepackage{amsfonts}
\usepackage{amssymb}
\usepackage[latin1]{inputenc}

\newcommand{\be}{\begin{equation}}
\newcommand{\ee}{\end{equation}}
\newcommand{\ben}{\begin{eqnarray}}
\newcommand{\een}{\end{eqnarray}}

\begin{document}
\title{Two Dimensional Horava-Lifshitz Black Hole Solutions}
\author{D. Bazeia$^{a}$, F.A. Brito$^{b}$ and F.G. Costa$^{a}$}
\affiliation{$^a$ Departamento de F\'\i sica, Universidade Federal da Para\'\i ba,
Caixa Postal 5008, 58051-970 Jo\~ao Pessoa, Para\'\i ba, Brazil \\
$^b$ Departamento de F\'\i sica, Universidade Federal de Campina Grande,
Caixa Postal 10071, 58109-970 Campina Grande, Para\'\i ba, Brazil}


\begin{abstract}
In this paper we address the issue of black hole solutions in (1+1)-dimensional non-projectable Horava-Lifshitz gravity. We consider several models by considering different potentials in the scalar matter sector.
We also consider the gravitational collapse of a distribution of pressureless dust filling a region in one-dimensional space. The time of the collapse can be faster or
slower depending on the parameter $\lambda$ of the theory. 
\end{abstract}
\maketitle



\section{Introduction}

The (1+1)-dimensional theory of gravity has been considered in some detail in several studies in the literature \cite{mann1,mann, sikmann, callan,fulling,trivedi}. Remarkably it has similarity to four-dimensional general relativity in many aspects. These include a Newtonian limit, cosmological solutions, interior solutions, gravitational waves and the gravitational collapse of pressureless dust into black holes with event horizon structures which is identical to four-dimensional Schwarzschild solution. Since from the classical point of view the (1+1)-dimensional gravity structure is so close to (3+1)-dimensional gravity it is expected that its quantization procedure should be quite similar to that in (3 + 1)-dimensional quantum gravity. Furthermore its semiclassical properties also produces interesting effects such as Hawking radiation and also black hole condensation. This is because the non-trivial  event horizon structure developed in (1+1)-dimensional theory of gravity has similarities with their (3 + 1)-dimensional general relativistic counterparts.

Recently in \cite{Horava:2009uw} it was put forward a new theory of gravity. This is now well-known as the Horava-Lifshitz (HL) gravity. In the HL gravity it is intended to obtain a renormalizable four-dimensional theory of gravity via power-counting due to higher order scaling on the three-momentum at UV scale. The price to pay is that space and time now scales in a different way via a dynamical critical exponent in the UV regime and as a consequence the Lorentz invariance is lost at high energy scale. Despite of this, several studies have been considered in the literature \cite{visser}, including modifications of the original theory in order to circumvent undesirable extra modes \cite{blas1}. However, to our knowledge, in low-dimensional HL gravity there have been few studies considered in the literature. To quote a few, very recently appeared considerations on black hole solutions in 2+1 dimensions  \cite{Park:2012ev, Sotiriou:2014gna,Lin:2014ija} and quantization of the (1+1)-dimensional projectable Horava-Lifshitz gravity \cite{wang}.

In the present study, we investigate black hole solutions and gravitational collapse of a pressureless dust distribution in 1+1 dimensions. We shall consider the non-projectable version of HL gravity \cite{blas1}. 

The paper is organized as follows. In Sec.~\ref{LD-HL-0} we briefly introduce our setup. In Sec.~\ref{LD-HL},  we are able to find several black hole solutions by considering distinct models through several specific potentials in the scalar matter sector. In Sec.~\ref{CG-HL} we shall assume only dust in the matter sector to address the issue of gravitational collapse. In Sec.~\ref{conclu} we present our final considerations.

\section{ Lowest Dimensional HL Theory}
\label{LD-HL-0}
In this section we shall briefly review the non-projectable HL gravity. In Horava-Lifshitz  gravity the spacetime decomposes as  follows
\be ds^2=-N^2dt^2+g_{ij}\left(dx^{i}+N^{i}dt\right)\left(dx^{j}+N^{j}dt\right).\ee
For this theory one finds
\be K_{ij}=\frac{1}{2N}(\dot{g}_{ij}-\nabla_{i}N_{j}-\nabla_{j}N_{i})\ee
and the $(D+1)$-dimensional action is written as
\be S=\frac{M^{2}_{Pl}}{2}\int d^{\textit{D}}xdt\sqrt{g}\left(K_{ij}K^{ij}+\lambda K^2+\cal{V}\right),\ee
where $\lambda>1$ and the potential ${\cal V}$ is associated with the non-projectable HL gravity defined as
\be {\cal V}=\xi R+\eta a^{i}a_{i}+\frac{1}{M^{2}_{\ast}}L_{4}+\frac{1}{M^{2}_{\ast}}L_{6},\ee
being $a_i$ a vector that describes the proper acceleration of the vector field of unit normals to the foliation surfaces \cite{blas1} given by
\be\label{a1-ln-N} 
a_{i}=\partial_{i}\ln N,
\ee
where $i=1,2,3$ in $3+1$ dimensions though we shall concentrate ourselves
in $1+1$ dimensions.
In $1+1$ dimensions the theory turns out to be much simpler, such that
\be K=\frac{1}{2N}(\dot{g}_{11}-2\nabla_{1}N_{1})\ee
with $z=D=1$, $g_{ij}=g_{11}$ and $a_{i}=a_{1}$. In this form, the HL action coupled to matter fields
\be\label{action0} S=S_{HL}+S_{\phi}\ee
becomes written in terms of 
\be\label{action1} S_{HL}=\frac{M^{2}_{Pl}}{2}\int dxdt\sqrt{g}\;[(1-\lambda)K^{2}+\eta g^{11}a_{1}a_{1}]\ee
and
\be S_{\phi}=\int dxdtN \sqrt{g}\;\left[\frac{1}{2N}(\partial_{t}\phi-N^{1}\nabla_{1})^2-\alpha(\nabla_{1}\phi)^2-V(\phi)-\beta\phi\nabla^{1}a_{1}-\gamma\phi a^{1}\nabla_{1}\phi\right]\ee
with $\alpha, \beta$ and $\gamma$ being dimensionless coupling constants. In the relativistic limit, we have $\alpha=1$ and $\beta=\gamma=0$ \cite{wang}.

\section{Lowest Dimensional Black Holes in HL Theory}
\label{LD-HL}

Notice that from Eq.~(\ref{action1}) the $\lambda$ parameter will be irrelevant for the static black hole solutions of the present section, because in this case $K=0$, whereas the $\eta$ parameter 
will be present in the most of the solutions. One should mention that in next section, where we shall address the problem of gravitational collapse, it will occur the opposite because 
in that case $a_1\to0$ (projectability).

Now rewriting the complete action using the fact that  $K=0$, by admitting $N_1(x)=0$ in our case, we have
\be\label{action} S=\frac{M^{2}_{Pl}}{2}\int dxdt\sqrt{-g}\left(\eta g^{11}a_{1}^2-\frac{2}{M^{2}_{Pl}}\alpha g^{11}\phi'^2-\frac{2}{M^{2}_{Pl}}V(\phi)\right),\ee
or simply 
\be\label{action2} S=\frac{M^{2}_{Pl}}{2}\int dxdt\left(-\eta N^2a_{1}^2+\frac{2}{M^{2}_{Pl}}\alpha N^{2}\phi'^2-\frac{2}{M^{2}_{Pl}}V(\phi)\right)\ee
since  $\sqrt{-g}=1$ in the present study --- see below.
Now varying this action with respect to $N$ it is easy to get the following important condition 
\be\label{cond-phi-a1} -\eta a_{1}^2+\frac{2}{M^{2}_{Pl}}\alpha \phi'^2=0,\ee
whereas by varying $S$ with respect to the scalar field $\phi$ we find
\be\label{EOM-phi}
\frac{d}{dx}(N^2\phi')=\frac{1}{2\alpha}\frac{\partial V}{\partial\phi}.
\ee
Notice from Eq.~(\ref{cond-phi-a1}) that  for $\alpha=\eta M_{Pl}^2/2$  we can find $a_{1}=|\phi'|$.
This gives us an important relation between the matter scalar field and $a_1$. From the non-projectable HL theory we can take advantage of Eq.~(\ref{a1-ln-N}), which in 1+1 dimensions
simply becomes 
\be\label{eq-for-N} a_{1}\equiv\frac{d\ln N}{d x}=|\phi'|\to N=e^{\pm\phi},\ee
where we have used the aforementioned condition between $a_1$ and $\phi'$ and integrated out the equation for $N$.

The equation of motion Eq.~(\ref{EOM-phi})  can be now written in the simpler form
\be\label{EOM-phi-2}
\frac{d}{dx}(\eta e^{2\phi}\phi')=V_\phi, \qquad V_\phi\to \frac{1}{M_{Pl}^2}\frac{\partial V}{\partial\phi}
\ee
where we used the above definition of $\alpha$ and the solution for $N$ given in Eq.~(\ref{eq-for-N}). The Planck mass appeared here because we kept $2/\kappa^2\equiv M_{Pl}^2/2$
in the Lagrangian (for later convenience --- see next section) although $[\kappa]\!=\!\frac{z-D}{2}\!=\!0$ since in our case $z\!=\!D\!=\!1$. Thus, for the moment we can indeed consider $M_{Pl}^2=1$ for simplicity and
consistence.

Let us now focus on the following model with $V(\phi)=0$. By using the equation of motion (\ref{EOM-phi}) we find
\be N^2\phi'=M,\ee
where $M$ is an integration constant. Now using Eq.~(\ref{eq-for-N}) we find
\be N\frac{dN}{dx}=\pm M,\ee
that integrating for $N(x)$ we find the general solution
\be N(x)=\sqrt{2\left(M|x|+C\right)}.\ee
For an integration constant chosen as $C=-1/2$  we find
\be N(x)^2=2M|x|-1.\ee
Consequently, the scalar field can also be found via relation $N=e^{\phi}$ given in Eq.~(\ref{eq-for-N}) such that
\be\label{sol-phi-0} \phi(x)=\ln\sqrt{2M|x|-1}.\ee 
This scalar solution can be thought of as a dilatonic solution. Its diverging behavior near the horizon is in accord with two other well-known places where the same phenomenon develops. Whereas our present study of black holes in low-dimensional  (two-dimensional)  Horava gravity has some resemblance with low-dimensional nonextremal black $Dp$-branes (e.g. $p<3$) in string theory \cite{bbs} the fact  that Horava gravity is in general a higher-derivative theory of gravity is somehow connected with higher-derivative gravity in arbitrary dimensions --- see for instance, the dilatonic Einstein-Gauss-Bonnet in Ref.~\cite{ohta-torii}. In these both situations the dilatonic solutions diverge on the black hole horizon. In the following we shall focus on the black hole solutions.

Thus, we obtain the following simplest solution of black hole in two-dimensional HL gravity
\be\label{sol-bh-0} ds^2=-\left(2M|x|-1\right)dt^2+\frac{1}{\left(2M|x|-1\right)}dx^2\ee
This solution has been previously appeared in \cite{mann1}.
Of course, since we have a scalar potential $V(\phi)$ that in general does not vanish, it is very natural to look for other solutions. However, as we shall see, it is not possible to find analytical solutions in some interesting cases. Despite of this, we shall consider the following models.

Firstly, let us consider the model with $V(\phi)=\Lambda \phi$.
By using the equation of motion (\ref{EOM-phi-2}) and the fact that $N=e^{\phi}$ we get the following equation
\be \eta(NN''+N'^2)-V_{\phi}=0.\ee
This equation can be solved analytically whose solution for $N(x)$ and $\phi(x)$ is given respectively by
\be N(x)^2=(\Lambda/\eta) x^2-2C_{1}x+2C_{2},\ee
and
\be \phi(x)=\ln \left[(\Lambda/\eta)  x^2-2C_{1}x+2C_{2} \right]^{1/2}.\ee
Now taking $C_{1}=-M$ and $\epsilon=2C_{2}$ we find
\be \phi(x)=\ln \left[(\Lambda/\eta)  x^2+2Mx-\epsilon\right]^{1/2}\ee
and also
\be\label{sol-phi-1} N(x)^2=(\Lambda/\eta)  x^2+2Mx-\epsilon.\ee 
Thus, in the present model, the new solution of black hole in two-dimensional HL gravity is
\be\label{sol-bh-1} ds^2=-\Big((\Lambda/\eta)  x^2+2Mx-\epsilon\Big)dt^2+\frac{1}{\Big((\Lambda/\eta) x^2+2Mx-\epsilon\Big)}dx^2.\ee
See in \cite{mann} (the first reference) a similar solution.
Before presenting more examples, some comments are in order. The cases studied previously are the simplest ones where we can choose a scalar potential and obtain explicit solutions. 
For further generalized potentials we cannot in general obtain explicit solutions analytically. In this sense, one can choose, instead, not a scalar potential itself but its derivative as a function of 
an implicit scalar field which in turns depends on the spatial coordinate. So we shall now consider forms of $V_\phi(\phi(x))\equiv V_\phi(x)$ as follows
\be\label{V(x)} V_{\phi}(x)=A+\frac{B}{x^3}+\frac{C}{x^4}.\ee
Now, substituting (\ref{V(x)}) into equation of motion (\ref{EOM-phi-2}) we find the following general solution for $N(x)$ and $\phi(x)$ given explicitly by
\be\label{sol-gen-bh} N(x)^2=2C_{2}+\frac{A}{\eta}x^2-2C_{1}x+\frac{B}{\eta x}+\frac{C}{3\eta x^2}\ee 
and
\be\label{sol-gen-phi} \phi(x)=\ln \sqrt{2C_{2}+\frac{A}{\eta}x^2-2C_{1}x+\frac{B}{\eta x}+\frac{C}{3\eta x^2}}.\ee
The only problem with this procedure is finding the potential back in terms of the scalar field $\phi$, because in most of cases one cannot invert the solutions in order to obtain $x=f(\phi)$.
Aside from  this fact, we can find several interesting solutions for $N(x)$ and $\phi(x)$ given explicitly, as we can see below. 
By properly choosing the parameters the spacetime may represent a black hole, a white hole, a naked singularity, or other more complicated structures. As stated in \cite{mann1}, this spacetime 
can also be used to easily extended it to multiple point sources.

Some special cases are in order:
\\

\noindent
$(i)$ For $C_{1}=-M$, $C_{2}=-1/2$, $\eta=1$ and $A=B=C=0$  we simply have $V_{\phi}=0$ ($V=const$), which is equivalent to the case with $V=0$ whose solution is given by Eqs.~(\ref{sol-phi-0})-(\ref{sol-bh-0}).
\\

\noindent
$(ii)$ Another similar example is $C_{2}=-\epsilon/2$, $C_{1}=-M$, $A=\Lambda$ and $B=C=0$ for what we have $V_{\phi}=A$, which is also equivalent to the case $V(\phi)=\Lambda\phi$. 
The solution for this case is given by Eqs.~(\ref{sol-phi-1})-(\ref{sol-bh-1}). If one wants to leave the solution in the same form presented in \cite{mann1}, one can still consider $\eta=1$ and $A=-\Lambda/2$.

In the following we shall consider two-dimensional Schwarzschild and Reissner-Nordstr\"om-like solutions. These types of solutions have previously appeared, e.g.,  in Refs.\cite{fulling} and \cite{trivedi}, 
respectively.

\subsection{Schwarzschild-like solution}
\noindent

In this case one considers $C_{2}=1/2$, $B=-2M$, $\eta=1$ and $A=C=C_{1}=0$. This leaves 
\be\label{pot-schw} V_{\phi}(x)=-\frac{2M}{x^3}.\ee
This gives the following explicit solution for $N(x)$ and $\phi(x)$ 
\be N(x)^2=1-\frac{2M}{x}\ee
and
\be \phi(x)=\ln\sqrt{1-\frac{2M}{x}}.\ee
Interestingly, in this case we can invert (\ref{pot-schw}) and integrate on $\phi$ to obtain the scalar potential
\be V(\phi)=\frac{B}{8M^3}\phi-\frac{3B}{16M^3}e^{2\phi}+\frac{3B}{32M^3}e^{4\phi}-\frac{B}{48M^3}e^{6\phi}.\ee
This will be quite easy anytime the polynomial form of $V_{\phi}(x)$ is restricted to a unique term.

\subsection{Reissner-Nordstr\"om-like solution}
\noindent
One can expect that the last term in $V_\phi(x)$ given in Eq.~(\ref{V(x)}) is associated with the effect of an electrical {\it charge} $Q$. This is more evident through the use of the general solution (\ref{sol-gen-bh})-(\ref{sol-gen-phi}) and making a suitable choice of the parameters, that is, for 
$C_{2}=1/2$,$B=-2M$, $C=3Q^2$, $\eta=1$ and $A=C_{1}=0$
\be\label{pot-schw2} V_{\phi}(x)=-\frac{2M}{x^3}+\frac{3Q^2}{x^4}.\ee
Again, as in the previous cases, this gives the following explicit solution for $N(x)$ and $\phi(x)$ 
\be N(x)^2=1-\frac{2M}{x}+\frac{Q^2}{x^2}\ee
and
\be \phi(x)=\ln\sqrt{1-\frac{2M}{x}+\frac{Q^2}{x^2}}.\ee
Differently of the previous case,  now one cannot easily invert (\ref{pot-schw}) and integrate on $\phi$ to obtain a scalar potential.

\subsection{A new black hole solution}
\noindent
The two cases presented above are well-known solutions in four-dimensions with several issues addressed such as number of horizons, Hawking temperature, entropy and so on. The other cases with $B=C=0$ are 
also well-known in 1+1 dimensional gravity. Thus, we do not need to say more about them. However, in the case that we will present here we shall consider the Hawking temperature.

Just for maintaining the usual notation, let us rename the general solution (\ref{sol-gen-bh}) as follows $f(x)\equiv N(x)^2$, i.e.,
\be f(x)=2C_{2}+\frac{A}{\eta}x^2-2C_{1}x+\frac{B}{\eta x}+\frac{C}{3\eta x^2}.\ee
For  $C_{1}\neq0$, $C_{2}\neq0$, $B\neq0$ and $A=C=0$ we have  
\be f(x)=2C_{2}-2C_{1}x+\frac{B}{\eta x}.\ee
This solution develops the following horizons 
\be x_{h}^{\pm}=\frac{C_{2}}{C_{1}}\pm\sqrt{\Delta},\qquad \Delta=\frac{C_{2}^2}{C_{1}^2}+\frac{2B}{\eta C_{1}}.\ee
As $\Delta=0$ they degenerate, i.e., $x_{h}^+=x_{h}^-$.

The Hawking temperature is given in terms of the outer ($x_h^+$) horizon as follows
\be T_{H}=\left.\frac{f'(x)}{4\pi}\right|_{x=x_{h}^+}.\ee
For the special case $C_{2}=0$, $C_{1}=-M$ and $B=-2M$ the horizons are independent of the mass $M$:
\be x_{h}^\pm=\pm\frac{2}{\sqrt{\eta}}\;\;(\eta>0)\ee
The temperature is then given by
\be T_{H}=\frac{1}{4\pi}\left(-2C_{1}-\frac{B}{(\frac{2}{\sqrt{\eta}})^2}\right),\ee
or simply
\be T_{H}=\frac{1}{8\pi}\left(4+\eta\right)M.\ee
This is a typical relation between the Hawking temperature and the mass of black holes in $1+1$ dimensions \cite{mann1}.

\section{Gravitational Collapse}
\label{CG-HL}

In this section we address the issue of the gravitational collapse of a certain mass of dust with negligible pressure confined into a region of the unidimensional space $[-r,r]$ which metric is given in co-moving 
coordinates by
\be\label{metric-colap} ds^2=-N(\tau)^2d\tau^2+a(\tau)^2d\rho^2\ee
In this case the action we shall consider is that given by Eq.~(\ref{action0}) with the matter sector not restricted only to scalar fields. Now we have the following action
\be\label{action1-1} S=\frac{M^{2}_{Pl}}{2}\int d^2xN\sqrt{g_{11}}\left[(1-\lambda) K^2+\eta g^{11}\phi'^2\right]+S_{m},\ee
that are explicitly given in terms of the metric (\ref{metric-colap}) in the form
\be S=\frac{M^{2}_{Pl}}{2}\int d^2x\left[\frac{(1-\lambda) a^3\dot{a}^2}{N}\right]+\frac{M^{2}_{Pl}}{2}\int d^2x\left[\frac{N\eta \phi'}{a}\right]+\int d^2xN\sqrt{g_{11}}L_{m}.\ee
The tensor energy-momentum is given in terms of the matter Lagrangian through its usual definition
\be T_{\mu\nu}=\frac{2}{\sqrt{-g}}\frac{\delta(\sqrt{-g}L_{m})}{\delta g_{\mu\nu}}.\ee
By varying the action with respect to $N$, i.e., 
\be \frac{\delta S}{\delta N}=0,\ee
we obtain the equation that relates the dynamics of the spacetime (\ref{metric-colap}) with the energy density
\be\frac{(\lambda-1) M^{2}_{Pl}a^3\dot{a}^2}{2N^2}+\frac{M^{2}_{Pl}\eta \phi'}{2a}=-\frac{1}{\sqrt{g_{11}}}\frac{\delta S_{m}}{\delta N}=\sigma.\ee
Now making $N=1$, and recalling that $N=e^\phi$ from our previous investigations, then consequently the scalar field $\phi=0$,  so that we have
\be\label{master-eq} (\lambda-1) M^{2}_{Pl}a^3\dot{a}^2=2\sigma.\ee

\subsection{Interior solution for the gravitational collapse}
\noindent
Since we are working with a pressureless fluid then $T^{\mu\nu}=\sigma U^\mu U^\nu$, being $U^1=0$ and $U^t=1$. Thus, the equation for conservation of energy and momentum now reads
\be \nabla_{\mu}T^{\mu\nu}=0\longrightarrow \frac{\partial}{\partial t}(\sigma\sqrt{a^2})=0,\ee
which means that $\sigma\sqrt{a^2}$ is constant.  Thus in terms of the constants $a_0$ and $\rho_0$ defined at initial time of the collapse it simply gives
\be \sigma=\frac{\rho_{0}a_{0}}{a}.\ee
Now substituting this equation into Eq.~(\ref{master-eq}) we find the following differential equation
\be a^4\dot{a}^2=\frac{2\sigma_{0}a_{0}}{M^{2}_{Pl}(\lambda-1)}\ee
that can still be recast in the form
\be a^2\dot{a}=\pm\sqrt{\frac{2\sigma_{0}a_{0}}{M^{2}_{Pl}(\lambda-1)}}=\pm\beta,\ee
whose solutions are
\be a=(\pm3\beta\tau+C)^{1/3}.\ee
Now choosing $C=1$, $N=1$ and taking the solution with minus sign we finally have the metric in the interior of the gravitational collapse 
\be ds^2=-d\tau^2+(1-3\beta\tau)^{2/3}d\rho^2.\ee
See in Ref.~\cite{sikmann} a similar solution.
The density of the dust given by $\sigma$ goes to infinity (singularity) as the scale factor approaches zero. This occurs in the finite time $\tau_c=1/(3\beta)$, that is
\be\tau_{c}=\sqrt{\frac{(\lambda-1) M^{2}_{Pl}}{18\sigma_{0}a_{0}}}.\ee
The square root dependence on $\lambda-1$ would be a problem in the projectable original HL gravity \cite{Horava:2009uw} where this parameter is allowed to be only $\lambda\leq1$. Fortunately this is not the case in the {\it healthy} non-projectable HL gravity developed in Ref.~\cite{blas1} where $\lambda>1$. Notice that given an initial density $\sigma_{0}$, the collapse can occur slower or faster depending on the parameter $\lambda$. As an example, for $\lambda\to 1$ the time $\tau_c\to0$, that means a distribution of dust 
that collapses very quickly, otherwise can live longer with a time $\tau_c\neq0$ before collapsing for $\lambda>1$.

\subsection{Exterior solution for the gravitational collapse}
\noindent
Inspired in the Birkhoff theorem, which states that is always possible to find a coordinate system in which the exterior solution of a spherical solution in 3+1 dimensions is time-independent  \cite{wein, Carroll:1997ar}, we shall proceed in a similar way into 1+1 dimensions to connect our {\it interior time-dependent} solution previously obtained to an {\it exterior time-independent} solution \cite{mann}.
Thus, we shall relate the coordinate $x$ that describes a black hole,  the static exterior solution, with a co-moving coordinate $\rho$ that describes the motion of the dust in the gravitational collapse, the interior solution, via 
\be x(\tau,\rho)=\rho a(\tau)=\rho(1-3\beta\tau)^{2/3}\ee
such a way that from the interior metric 
\be\label{comov} ds^2=-d\tau^2+a^2(\tau,\rho)d\rho^2,\ee
we should find the static exterior solution
\be\label{schw} ds^2=-A(x)^2dt^2+A(x)^{-2}dx^2.\ee
There is a Killing vector that corresponds to energy conservation  satisfying
\be K_{\mu}\frac{dx^{\mu}}{d\tau}=K_{t}\frac{dt}{d\tau}+K_{x}\frac{dx}{d\tau}=const,\ee
where
\be K_{\mu}=\left(-A^2,0\right).\ee
Then we find the following solution
\be\label{DT}\frac{dt}{d\tau}=-\frac{E}{A^2}\ee
In addition, there is another constant of the motion along the geodesic
\be \epsilon=-g_{\mu\nu}\frac{dx^{\mu}}{d\tau}\frac{dx^{\nu}}{d\tau},\ee
that is
\be -\epsilon=-A^2\left(\frac{dt}{d\tau}\right)^{2}+A^{-2}\left(\frac{dx}{d\tau}\right)^{2},\ee
or simply 
\be -\epsilon A^2=-A^4\left(\frac{dt}{d\tau}\right)^{2}+\left(\frac{dx}{d\tau}\right)^{2},\ee
\be\left(\frac{dx}{d\tau}\right)^{2}+\epsilon A^2-E^2=0.\ee
If the particle is at rest  ($dx/d\tau=0$) in the limit $x\rightarrow r$ we have  $\epsilon E^2=C^2$ with $C\to A(r)$. Recall that $r$ is the boundary of the dust in a one-dimensional region. In the case of massive particles we can make  $\epsilon=1$ then
\be\left(\frac{dx}{d\tau}\right)^{2}+A^2-C^2=0.\ee
Notice also that from Eq.~\ref{DT} we can now write
\be\frac{dt}{d\tau}=-\frac{C}{A^2}\ee
Imposing the conditions $x=x(\rho,\tau)$, $t=t(\rho,\tau)$ and $x=\rho a$ we have
\be dx=\frac{\partial x}{\partial \tau}d\tau+\frac{\partial x}{\partial \rho}d\rho=\rho\frac{\partial a}{\partial \tau}d\tau+ad\rho\ee
\be dt=\frac{\partial t}{\partial \tau}d\tau+\frac{\partial t}{\partial \rho}d\rho=-\frac{C}{A^2}d\tau+\frac{\partial t}{\partial \rho}d\rho \ee
Substituting this into (\ref{schw}) and comparing with (\ref{comov}) we have the following conditions
\be\frac{\partial t}{\partial\rho}=\frac{1}{C^2}\frac{1}{A^2}\rho a \frac{\partial a}{\partial\tau}=\frac{1}{C}\frac{1}{A^2}\rho\frac{\beta}{a}\ee
\be A^2=C^2-\rho^2\left(\frac{\partial a}{\partial \tau} \right)^2=C^2-\frac{\beta^2}{a^4}\rho^2\ee
\be \left(\frac{\partial t}{\partial \rho}\right)^2=\frac{a^2}{A^4}-\frac{a^2}{A^2}\ee
From where we conclude that 
\be A^2=1-\frac{\beta^2}{a^4}\rho^2\ee
Making $\rho=r$ and $x=ra$, we have
\be A^2=1-\frac{\beta^2r^6}{x^4}\ee
\be ds^2=-\left(1-\frac{\beta^2r^6}{x^4}\right)dt^2+\left(1-\frac{\beta^2r^6}{x^4}\right)^{-1}dx^2\ee
This is precisely one of the solutions found in  \cite{sikmann}.
The scalar curvature is given by
\be R=\frac{20\beta^2r^6}{x^6}\ee
Thus  $x=0$ is a truly singularity, that is, a curvature singularity. Furthermore, the  two-dimensional Schwarzschild radius is given by
\be x_{H}=r^{3/2}\beta^{1/2}=r^{3/2}\left[\frac{2\sigma_{0}a_{0}}{M^{2}_{Pl}(\lambda-1)}\right]^{1/4}.\ee
The previous analysis has many similarities with that considered long ago \cite{sikmann}. However, there are some peculiar points here. It seems the most striking difference is the one related to the  Schwarzschild radius. In the (1+1)-dimensional Einstein gravity explored in \cite{sikmann} the Schwarzschild radius is not defined at $r=1/\sqrt{4b}$, where $b=2\pi G\rho_0$ and $r$ is the boundary of the dust. On the other hand, the Schwarzschild radius in the present case is well defined for $\lambda>1$, which is also naturally consistent for a collapse at finite time discussed above.

\section{Conclusions}
\label{conclu}
We have investigate black hole solutions in the two-dimensional HL gravity. The solutions are in principle the same obtained in 1+1 general relativity but are controlled by the parameter $\eta$ which controls the coupling of the vector associated with the non-projectability of the theory. However, they do not depend on the coupling $\lambda>1$. The opposite happens to the gravitational collapse of the pressureless dust. In this case there is  no dependence on $\eta$ but it depends on $\lambda$. This is due to the specific dependence of the solutions on the coordinates in each case. Whereas the black hole solutions are only spatial dependent, the obtained {\it interior} solution for the gravitational collapse of dust has only time dependence. The exterior solution is just obtained from the interior solution.

\section{ADDENDUM}

By varying the action (\ref{action}) with respect to $N$ and $\phi$ we find the following equations
\ben
&&\frac{\delta S}{\delta N}=0\to-\eta\frac{N''}{N'}+\frac{2\alpha}{M_{Pl}^2}{\phi'}^2=0,\label{eqA-1}\\
&&\frac{\delta S}{\delta \phi}=0\to\frac{d}{dx}(N^2\phi')=\frac{1}{2\alpha}\frac{\partial V}{\partial\phi}\label{eqA-2}.
\een
For dilatonic-type solutions $N=e^{\phi}$, $N'=\phi'N$ and $N''=(\phi''+{\phi'}^2)N$.  Now assuming ``slow-varying'' dilatonic field, i.e., $\phi''\ll{\phi'}^2$, and recalling that $a_{i}=\partial_{i}\ln N\equiv \frac{N'}{N}$, we have that $N''/N={\phi'}^2\equiv a_1^2$, and the equation (\ref{eqA-1}) now reads
\ben
-\eta{a_1}^2+\frac{2\alpha}{M_{Pl}^2}{\phi'}^2=0,\label{eqA-3}
\een
which is our previous equation (\ref{cond-phi-a1}). The aforementioned slow-varying dilatonic field can be put in the following way:
\ben
\frac{\phi''}{{\phi'}^2}=\left(\frac{N''}{N}-\frac{{N'}^2}{N^2}\right)\frac{N^2}{{N'}^2}\ll1.
\een
This precisely happens in the {\it near horizon regime} where $N(x\!\to\! x_h)^2\!\to\!0$.

In conclusion, for the main purposes, we can ensure that the general solution (\ref{sol-gen-bh})-(\ref{sol-gen-phi}) is valid around black hole horizons.

\acknowledgments

We would like to thank to CNPq, PNPD-CAPES and PROCAD-NF/2009-CAPES for partial financial support.

\end{document}